%% file: main_tcss.tex
\documentclass[lettersize,journal]{IEEEtran}

\input{macros}

\begin{document}

\title{Variational Inference of Parameters in Opinion Dynamics Models}

\author{
\IEEEauthorblockN{Jacopo Lenti\IEEEauthorrefmark{1}\IEEEauthorrefmark{2}, Fabrizio Silvestri\IEEEauthorrefmark{1}, \texorpdfstring{\nohyphens{Gianmarco~De~Francisci~Morales}}{Gianmarco De Francisci Morales}\IEEEauthorrefmark{2}}\\
\IEEEauthorblockA{\IEEEauthorrefmark{1}Sapienza University, Rome, Italy.}
\IEEEauthorblockA{\IEEEauthorrefmark{2}CENTAI, Turin, Italy.}\\
\textit{jcp.lenti@gmail.com, fsilvestri@diag.uniroma1.it, gdfm@acm.org}
}






\maketitle

\begin{abstract}

Understanding human behavior represents a paramount challenge in modern social systems.
This task must be tackled with tools that both explain the mechanisms underlying the social dynamics and efficiently handle vast amounts of data.
While Agent-Based Models (ABMs) are generally used as simulating tools to describe social dynamics, their connection to data is lackluster.
Instead, here we adopt a probabilistic machine learning approach for fitting ABMs to real data.
To this end, we propose a Variational Inference (VI) framework that estimates the macroscopic and microscopic parameters of several opinion dynamics models.
Our methodology encompasses 3 steps: ($i$) translation of the opinion dynamics models into Probabilistic Generative Models (PGMs), ($ii$) relaxation of discrete variables to make the models differentiable, and ($iii$) estimation of the parameters and latent variables via Stochastic Variational Inference.
Experiments show that VI improves over existing methods in estimating discrete and continuous variables, both at the microscopic and macroscopic scales, in all 4 different categories of rules opinion dynamics models.
Moreover, VI effectively estimates high-dimensional variables, up to 400 agent-level attributes, and is faster than the alternatives.

\end{abstract}


\section{Introduction}

The rise of online social networks has induced a shift of human actions and interactions in the digital sphere~\cite{lorenz2019accelerating, bakshy2012role}.
This transformation has not only altered the dynamics of interaction and opinion formation,  but has also generated vast amounts of data, offering unprecedented opportunities to study human behavior and its societal implications~\cite{peralta2022opinion}.
For these reasons, investigating social data has become a critical area of research.
It has the potential to provide new insights into the mechanisms driving social phenomena.
To this end, the only way to understand human society from data is with interpretable models capable of both explaining causal effects and representing the collected data.
In this context, Agent-Based Models (ABMs) represent a viable path to provide a causal connection between individual actions and global properties of social systems~\cite{axtell2022agent}.

ABMs are computational frameworks that describe human behavior and the consequent emergence of complex phenomena.
They do so by simulating the actions and interactions of autonomous agents within an environment (e.g., a social network)~\cite{grimm2013individual}.
These models are popular across diverse scientific domains, such as biology, ecology, economics, and the social sciences~\cite{an2023modeling, grimm2013individual, stillman2015making, will2020combining}.
Indeed, a large portion of Computational Social Science literature employs ABMs to investigate the online dynamics driving the spread of misinformation~\cite{cinelli2021dynamics}, the formation of echo chambers~\cite{cinelli2021echo,minici2022cascade}, and health communication~\cite{prieto2021vaccination, sobkowicz2021agent}, to name a few.
In particular, opinion dynamics study how individuals adjust their beliefs and opinions as a result of peer interactions and media exposure~\cite{coates2018unified}, and are usually formulated as ABMs.

\begin{figure}[t]
\centering
\includegraphics[width=\linewidth]{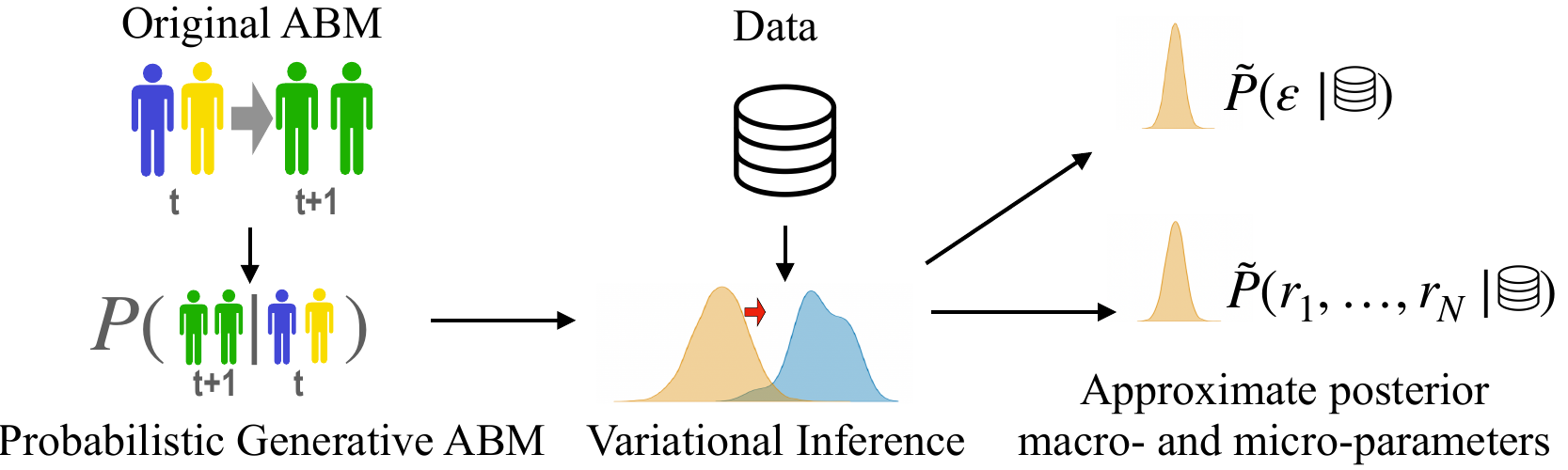}
\caption{The proposed parameter estimation pipeline. 
First, we translate the agent-based model into a probabilistic generative agent-based model. 
Then, we apply variational inference to get an approximate posterior of the target parameters within a given dataset.
}
\label{fig:diagram_pgabms}
\end{figure}

The mechanistic nature of the ABMs enables the exploration of what-if scenarios and counterfactual reasoning~\cite{casini2016agent}.
This capacity to simulate and analyze hypothetical situations, combined with the inherent flexibility of ABMs, has contributed to their rise in several disciplines.
However, despite their potential, the practical utility of ABMs is currently limited by the lack of connection between models and data~\cite{daly2022quo}.

Indeed, the heuristics underlying the theoretical rules often lack robust empirical validation against real-world data~\cite{casini2016agent}.
Aligning these models with real data through parameter tuning remains a significant challenge~\citep{flache2017models}.
Several factors contribute to these difficulties, including the considerable computational effort required to simulate the models and the high dimensionality of microscopic parameters, which scales with the size of the system.
Additionally, the heterogeneity of agents and the diversity of the rules in the ABM literature hinder the tractability of these models~\citep{platt2020comparison}.
Current practices for parameter estimation in ABMs predominantly rely on simulation-based methods.
These methods involve running a large number of computationally-expensive simulations of the model with varying parameter values and comparing the observed and simulated trajectories of the system state (a process called \emph{calibration})~\citep{fagiolo2019validation,lux2018empirical}.
Given the complexity of these models, the research community has focused on using machine learning and AI to enhance ABMs~\cite{an2021challenges, an2023modeling, lavin2021simulation}.

In this work, we infer the parameters of ABMs by adopting a pipeline borrowed from the probabilistic machine learning literature.
Rather than treating the models as black-box data generators, we exploit the rules of the ABMs to define Probabilistic Generative ABMs (PGABMs) and represent them within a Probabilistic Programming Language (PPL).
A PPL is a stochastic simulator which uses forward simulations to infer the parameters based on the observed outcome of the model~\cite{lavin2021simulation}.
This approach involves translating the ABM rules into conditional probabilities, inherently incorporating uncertainty into the models.
Thus, we provide a methodology that learns the parameters from the data, considering the ABM as the data-generating process, thus showing the applicability of probabilistic machine learning for connecting ABMs to data.

Specifically, we propose a pipeline to infer the parameters with Variational Inference (VI).
VI assumes a tractable, parametric, functional form for the approximation of the posterior distribution of the target parameters.
Then, it minimizes the \corr{Kullback-Leibler (KL)} divergence between the approximated distribution and the real posterior.
To do so, we need to first make the PGABM differentiable to then optimize the variational distribution.
In so doing, our VI approach can address models with intractable likelihood functions (e.g., with a large number of categorical variables).
\Cref{fig:diagram_pgabms} depicts the proposed pipeline.

To show the validity of our proposal, we tackle parameter estimation in 4 different variants of the Bounded Confidence Model (also known as the Deffuant-Weisbuch model), one of the most popular models in opinion dynamics~\cite{weisbuch2002meet, lorenz2007continuous}.
We show that our method can estimate both agent-level attributes (\emph{microscopic} parameters) and system-level parameters (\emph{macroscopic} parameters), both continuous and discrete.
In our experiments, VI improves the average Root-Mean-Square Error (RMSE) of macroscopic variables by a factor ranging between 2.5 and 12.1 compared to a simulation-based method (Approximate Bayesian Computation).
Moreover, our approach is up to 50 times faster, and it can recover up to $400$ binary microscopic agent attributes, a task that is unfeasible in traditional simulation-based inference (\corr{which would require visiting $2^{400}$ of parameters}).

The proposed technique offers several important advantages, compared to simulation-based methods.
\begin{squishlist}
        \item PGABMs are based on the theory of \corr{Probabilistic Generative Models (PGMs)}, which offers a common ground toward a unifying formalism for ABMs.
        \item VI tackles the inference problem as an optimization task, estimating the parameters without repeatedly simulating the model, thus alleviating its computational cost.
        \item Simulation-based methods ignore information not captured by the summary statistics.
        Instead, our approach leverages the full information contained in the original data.
        This significantly limits issues of equifinality, where multiple model parameterizations lead to the same outcome, and produces empirically more accurate estimates.
\end{squishlist}

\section{Related Works}
Most of the current methodologies for estimating parameters in ABMs are simulation-based.
Despite the many technical differences between the methods, all simulation-based approaches share some common steps: (i) a sampler draws the parameters from the parameter space, (ii) a set of simulations is run with the selected parameters, (iii) the simulated data trajectories are reduced to low-dimensional summary statistics, and (iv) the simulated and observed summary statistics are compared to assess the quality of the parameters in the fitness of the model~\citep{fagiolo2019validation}.
Hence, the results of the estimation depend on some arbitrary choices: the summary statistics to use and the notion of distance between them~\citep{hartig2011statistical}.
Moreover, the computational cost required by the simulations makes an exhaustive search of the parameter space unfeasible in most cases.

Among simulation-based methods, Simulated Minimum Distance (SMD) methods are often employed. 
These methodologies follow the simulation-based routine described before to select the set of parameters that best replicate some desired properties~\cite{franke2009applying, glielmo2023reinforcement}.
In a Bayesian setting, Approximate Bayesian Computation (ABC) represents the cornerstone of simulation-based inference~\cite{cranmer2020frontier,carrella2021no,grazzini2017bayesian}.
ABC selects the parameters from the prior and then runs the simulations.
Therefore, the sampled parameters are accepted or rejected to estimate the posterior.

Remarkable progresses in inference on ABMs addresses the reduction of the simulation burden, by effectively exploring the parameter space~\cite{grazzini2017bayesian}, and by efficiently generating simulated data traces with machine learning surrogates~\cite{lamperti2018agent}.
In particular, by relying on Differentiable ABMs, it is possible to employ Automatic Differentiation (AD) in the inference phase.
This requires approximating the categorical distributions with continuous relaxations.
In this way, it is possible to compute the gradient of the loss during the simulations and explore the parameter space with gradient descent~\cite{andelfinger2021differentiable, quera2023bayesian, dyer2022black, dyer2023gradient}.

Different approaches translate the ABMs into state space models, applying particle filters (i.e., Sequential Monte Carlo)~\cite{lux2018estimation}.
Similarly, Hidden Markov Models (HMMs) can represent ABMs as state space models~\cite{dong2016variational, fang2017expectation}.
By using mean-field approximations, these last works employ a variational approach to estimate agents' state transition probabilities~\cite{dong2016variational, fang2017expectation}.

A recent line of work has suggested tackling the parameter estimation problem in ABMs radically differently: through a likelihood-based approach~\citep{monti2020learning,monti2022learning,lenti2024likelihood}.
These works propose a paradigm shift by translating the ABMs into probabilistic generative models, called Probabilistic Generative ABMs (PGABMs).
PGABMs explicitly define the latent and observed variables of the system according to the available data and describe the conditional probabilities connecting them to the rules of the ABMs.
This representation allows for the derivation of the likelihood function of the latent variables from the observed data, especially by leveraging the theory of PGMs to extrapolate the conditional independencies among the variables at play.
As a result, this approach is more principled, more accurate, and faster than simulation-based alternatives~\citep{lenti2024likelihood}.
Nevertheless, the likelihood-based approach requires a nontrivial analytical process to derive the likelihood function.
In this work, we adopt the PGABMs framework to cast the model into a PPL.
In this way, we estimate the parameters during the forward simulations, instead of minimizing the negative log-likelihood.
This permits to avoid the step of derivation of the likelihood, which is considered unfeasible in the majority of ABMs.

\section{Preliminaries}
\label{sec:prelim}
\subsection{Nomenclature}
\label{sec:nomenclature}
In the literature, the abbreviation PGM is used both to refer to ``Probabilistic Generative Model'' and ``Probabilistic Graphical Model''.
\corr{In this paper, PGM refers to ``Probabilistic Generative Model'', while PGABM is ``Probabilistic Generative ABM''.
Probabilistic Graphical Models are graphical representations of PGMs via direct acyclic graphs.}

In the field of ABMs, and dynamical systems in general, a \emph{variable} depends on time, while a \emph{parameter} is independent of time.
Conversely, the nomenclature for PGMs is more concerned with randomness:
a variable can assume different values according to either a stochastic or deterministic process.
An observed variable without parents in the PGM is called a \emph{parameter}.
It should be clear from the context when we are referring to a latent ABM parameter that we are estimating vs. an observed PGM parameter that is given.
Similarly, the same objects are treated as ABM \emph{parameters} and PGM \emph{latent variables}, and this distinction should be clear from the context.

\subsection{Variational Inference}
\label{sec:vi}
In the typical Bayesian framework, the objective is to determine the posterior distribution $p(\theta \mid y)$ of the model parameters $\theta$ given the observed data $y$.
Bayes' theorem provides the well-known relationship $p(\theta \mid y) =  p(y \mid \theta) p( \theta) / p(y)$, where $p(\theta)$ is the prior distribution of the model parameters which encodes our initial beliefs about $\theta$, and $p(y \mid \theta)$ is the likelihood of observing the data according to the underlying probabilistic model.
However, as the evidence $p(y) = \int p(\theta, y) \, d\theta$ is generally intractable, closed-form solutions for $p(\theta \mid y)$ are limited, thus Bayesian inference often relies on approximations.

\mcmc is one of the most common techniques: it uses a sampling routine for $\theta$ based on the prior distribution and is guaranteed to converge to the correct posterior asymptotically.
However, \mcmc algorithms do not scale well to large datasets and often struggle to approximate multimodal posteriors~\citep{blei2017variational}.

Variational Inference (VI) is an alternative to MCMC for approximating a target density via optimization~\citep{blei2017variational}.
VI approximates the target posterior $p(\theta \mid y)$ by using a more tractable family of densities $q_\lambda(\theta)$, called \emph{variational distribution}, indexed by the variational parameter $\lambda$.
The optimal variational parameter $\lambda^*$ is the one that maximizes the evidence lower bound (ELBO), which is a lower bound of the likelihood.
The ELBO is defined as 
\begin{align}
    \mathcal{L}(\lambda) = \mathbbm{E}_q [\log p(y, \theta) - \log q_\lambda (\theta)].
    \label{eq:ELBO}
\end{align}
By writing 
\begin{align}
\mathcal{L}(\lambda) = \mathbbm{E}_q [\log p(\theta \mid y) + \log p(y) - \log q_\lambda (\theta)],
\end{align}
we obtain
\begin{align}
\mathcal{L}(\lambda) = \log p(y) - \mathbbm{E}_q [\log q_\lambda (\theta) - \log p(\theta \mid y)].
\end{align}
The second term of the right-hand side of this equation is $KL(q_\lambda(\theta) \;\|\; p(\theta \mid y) )$.
Thus, maximizing the ELBO with respect to $\theta$ is equivalent to minimizing the KL-divergence of  $q_\lambda(\theta)$ from $p(\theta \mid y)$ (up to a constant).
In this way, we tackle the task of inferring the most fitting parameters as an optimization problem (i.e., maximization of the ELBO), improving the efficiency as the dataset size increases.

\spara{Stochastic Variational Inference.}
Stochastic Variational Inference (SVI) is a common routine for maximizing the ELBO~\citep{hoffman2013stochastic}.
SVI uses gradient ascent to optimize the ELBO and estimates the gradient on subsamples of the data instead of using the entire dataset for improved speed.

\spara{Variational Inference with Normalizing Flows.}
The choice of the variational distribution, $q_\lambda$, represents a critical step in the VI process, as it encodes the functional form of the approximation of $p(\theta \mid y)$.
The family of normal distributions is a common choice thanks to its simplicity and prevalence in Bayesian statistics.
However, the normal family imposes strong constraints on the variational distribution, such as symmetry and unimodality.
A recent and more flexible solution is to adopt normalizing flows (NFs) as the variational distribution~\citep{rezende2015variational}.
NFs are transformations of simple probability distributions (e.g., normal) into richer and more complex distributions through a sequence of invertible and differentiable mappings.
The expressiveness of these transformations enables NFs to represent arbitrarily complex distributions, from which we can efficiently sample and compute the ELBO~\citep{papamakarios2021normalizing}.

\spara{Categorical variables.}
Maximizing the ELBO via gradient ascent is feasible only for continuous, differentiable variables.
Categorical variables pose a great challenge, both because of their discrete nature and because there is no inherent ordering within their support.
A common approach is to replace the discrete variables with continuous, tractable approximations via the reparameterization trick.
The Gumbel-Softmax reparameterization offers a viable solution for these cases~\citep{maddison2016concrete,jang2016categorical}.
Let $X$ be a categorical random variable with $P(X=k)\propto\alpha_k$ and let $\{G_k\}_{k<K}$ be an i.i.d. sequence of Gumbel random variables.
Then, $X = \arg\max_k(\log \alpha_k + G_k)$.
The $\arg\max$ function is thus replaced by the softmax with temperature $\tau$,
\begin{align*}
    X_k = \frac{\exp(\log \alpha_k + G_k) / \tau }{\sum_{i=1}^K \exp(\log \alpha_i + G_i) / \tau} .
\end{align*}

A temperature $\tau$ parameterizes the softmax and controls how closely it approximates the argmax (closer for $\tau \to 0$).
Consequently, we obtain a continuous variable of which we can compute the gradient and learn $\alpha_k$ with an automatic differentiation (AD) routine.
In our experiments we use $\tau = 0.1$.
If $k = 2$, the Gumbel-Softmax reparameterizes a Bernoulli distribution.
This technique provides a way to relax a probabilistic generative model with categorical random variables, which allows for optimizing the ELBO and, consequently, estimating the posterior distribution of the target parameters.


\section{\corr{Models Overview}}
\label{sec:overview}
In the following sections, we show that our VI approach can successfully estimate the parameters of a variety of opinion dynamics models.
Specifically, we start from a Bounded Confidence Model with backfire effect (BCM-b)~\citep{jager2005uniformity}, and we extend it in four different directions.
Thanks to its ability to replicate several macroscopic properties---consensus, fragmentation, and polarization---from a few simple microscopic behaviors, BCM-b represents a cornerstone in the opinion dynamics literature~\cite{noorazar2020classical}.

BCM-b illustrates the dynamics in a social network with agreement and disagreement behaviors.
Each agent has an opinion in $[0,1]$, where 0 and 1 represent the two polar opposites on a given topic (e.g., on the left--right political spectrum).
At each interaction, an agent $u$ interacts with one of its neighbors $v$, thus emulating a social network.
If the opinions of the two interacting agents are closer than a threshold---the bounded confidence interval ($\eps^+$)---they have a positive interaction ($s = 1$), and their opinions converge by a convergence rate ($\mu^+$).
Conversely, if their opinions are further than a backfire threshold ($\eps^-$), then they have a negative interaction ($s = -1$), and their opinions diverge by a divergence rate ($\mu^-$).
In the other cases, their opinions remain unchanged ($s = 0$).
In our settings, an interaction $(u,v)$ is asymmetric since only the opinion of $v$ gets updated.
However, the same estimation methods are applicable in symmetric models.

Let $x^t_u$ be the opinion of agent $u$ at time $t$, the dynamics of the opinion after the interaction $e_t = (u,v)$ are as follows
\begin{align}
    \left\{
    \begin{array}{lcl}    
    x^{t+1}_v = x^{t}_v + \mu^+ \Delta x^t_{uv} & \text{if} & \lvert \Delta x^t_{uv} \rvert \leq \eps^+ ,\\
    x^{t+1}_v = x^{t}_v  & \text{if} & \eps^+ < \lvert \Delta x^t_{uv} \rvert < \eps^- ,\\
    x^{t+1}_v = x^{t}_v - \mu^- \Delta x^t_{uv} & \text{if} & \lvert \Delta x^t_{uv} \rvert \geq \eps^- ,\\
    \end{array}
    \right. 
    \label{BCMdet}
\end{align}
where $\Delta x^t_{uv}$ is $x^t_u - x^t_v$, and where $\eps^+ < \eps^-$.
After each update, the opinions are clamped to $[0,1]$, 
\begin{align}
    x^{t+1}_v = \max (0, \min ( x^{t+1}_v, 1) ) .
\end{align}

Different choices of $\eps^+$ and $\eps^-$ lead to different final configurations of the opinions at the end of the simulation (time step $T$), which can capture consensus, polarization, or fragmentation~\citep{jager2005uniformity}.
In our experiments, we constrain $\eps^+ \in [0,0.5]$ and $\eps^- \in [0.5,1]$ in order to avoid rapid opinion convergence.

\corr{Relying on a taxonomy that divides the opinion dynamics models into four categories~\cite{coates2018unified}, we extend BCM-b in different directions.}
We choose one representative rule per category, and estimate a set of parameters driving such rule:
\begin{squishlist}
\item \textbf{BCM-S}. \emph{Structural rules} comprise the initial setting of the system before running the simulations, such as the distribution of agents' attributes or the structure of the social network.
We estimate a binary attribute of the agents.
\item \textbf{BCM-I}. \emph{Interaction rules} govern who interacts with whom at each time step. Usually, agents can have synchronous or asynchronous interactions, they can be symmetric or asymmetric.
We estimate the number of influencing neighbors in a setting of higher-order interactions.\footnote{\corr{\citet{coates2018unified} call these ``Communication rules''.}}
\item \textbf{BCM-U}. \emph{Update rules} encompass the evolution of the opinions as a result of the interactions of the agents.
We estimate the presence of the backfire effect in the system, represented as an update rule that allows opinion divergence when the interacting opinions are far.
\item \textbf{BCM-G}. \emph{Graph rules} span the changes in the network structure as a consequence of agents' interactions, such as link formation and link rewiring.
We estimate the probability of link rewiring.\footnote{\corr{\citet{coates2018unified} call them ``Coevolutionary rules''.}}
\end{squishlist}

\section{Inference}
In our experiments, we estimate \eps and the parameters related to the specific rules of the model.
\corr{We focus on the parameters that govern the interaction outcomes, while leaving for future research the study of speed of convergence parameters ($\mu$), which is known to lead to biased maximum likelihood estimators~\citep{borile2025bias}}.
The general pipeline encompasses the following steps:
\begin{enumerate}
    \item Translate the rules of the ABM to derive the corresponding PGABM;
    \item Relax the discrete variables to make the PGABM differentiable;
    \item Estimate the latent variables of the PGABMs by maximizing the ELBO;
    \item Compare the estimates against the baselines.
\end{enumerate}

We emphasize that the goal of the paper is to provide an appropriate method for parameter inference.
To achieve this, we evaluate the capability of the methods to recover the correct parameters of a known data-generating process.
Consequently, we test our method via controlled experiments using synthetic data.
Since we need to know the generating process, using real data for this purpose would be unfeasible.

\spara{PGABM.}
Since all the models we analyze are extensions of BCM-b, some steps are common to all our experiments.
To translate the ABM into the corresponding PGABM, we need to define the variables of the PGM.
These variables can be observed, latent, or parameters, depending on data availability.
Then, we express the rules of the ABM as conditional probabilities.

In BCM-b, the outcome of the interaction $s$ is observed, and it can be positive, neutral, or negative.
To represent it, we define the binary variables $s^+$ and $s^-$, as indicator functions of $s$ being positive or negative, respectively.
The pair of interacting agents is $e$, a parameter of the PGM since it is observed and has no parents.
Similarly, the update parameter $\mu = (\mu^+, \mu^-)$ is another parameter of the PGM.
The variable $x_t \in [0,1]^N$ represents the opinion of the agents, where $N$ is the number of agents.
As we observe $x_0$, $x_t$ can be computed as deterministic functions of observed variables and parameters.
Finally, the target variable $\eps = (\eps^+, \eps^-)$ is latent.

\input{pgms_BCM}

Now, we define the conditional probabilities of observing the interaction outcomes $s^+$ and $s^-$
\begin{align}
    & P(s^+_j = 1 \mid x_t, e_j, \eps) = \sigma\left(\rho \cdot \left( \eps^+ - \lvert \Delta x^t_{uv} \rvert\right)\right)\label{eq:kappa_plus_BCMb}\\
    & P(s^-_j = 1 \mid x_t, e_j, \eps) = \sigma\left(-\rho \cdot \left(\eps^- - \lvert \Delta x^t_{uv} \rvert\right)\right)\label{eq:kappa_minus_BCMb},
\end{align}
where $\sigma(\cdot)$ is the sigmoid function and $\rho$ is its steepness~\citep{monti2020learning}.
This way, the probabilistic model generates $s^+ = 1$ with a high probability if $\lvert \Delta x^t_{uv} \rvert < \eps^+_{r_v}$, and $s^- = 1$ if $\lvert \Delta x^t_{uv} \rvert \geq \eps^-_{r_v}$.
This process ensures that any sample from the parameter space can be associated with a non-zero probability of having generated the data. 
As $\rho \rightarrow +\infty$ such probabilities approach $1$ and the model converges to the one in \Cref{BCMdet}.
In our experiments, we set $\rho$ = 32, for which the probabilities of interactions outside the confidence bounds are already negligible ($\leq 10^{-5}$).
Increasing $\rho$ leads to numerical problems, such as vanishing gradients, while decreasing $\rho$ provides less accurate approximations of \Cref{BCMdet}.
Note that the samples of $s^+$ and $s^-$ are independent, so each interaction can have both a positive and negative outcome at the same time, which results in a simultaneous convergence and divergence of the opinions.
This minor divergence from the original model is made for covenience.

After having derived the PGM, we need to infer the target latent variables.
We compare 4 different inference methods to do so:
two methods based on VI that represent our proposal,
an MCMC method that is a more classical take on the concept of PGABM,
and a Bayesian simulation-based method.

\spara{Variational Inference.}
The two methods based on VI differ in the family for the variational distribution.
One uses a multivariate normal distribution while the other uses Normalizing Flows.
The variational distribution $q_\lambda$ has dimension $M$, that is, the dimension of the estimand (which varies across the models).

Similarly to above, we relax each PGABM to make it differentiable.
The approximations of the discrete variables depend on the model and we present them in the specific subsections.
Once the model is differentiable, we cast it into a probabilistic programming language for the inference (NumPyro in our case~\cite{phan2019composable}).

Upon sampling $\theta \sim q_\lambda$, we need to restrict $\theta$ in the appropriate domains since $q_\lambda$ has support in $\mathbbm{R}^{M}$.
We compute $\hat{\eps}$ from $\theta_1$ and $\theta_2$ as follows,
\begin{squishlist}
    \item $\heps^+ = \sfrac{\sigma(\theta_1)}{2}$
    \item $\heps^- = \sfrac{\sigma(\theta_2)}{2} + \sfrac{1}{2}$
\end{squishlist}
\noindent Note that $\sigma(\cdot)$ transforms the domain from $(-\infty, +\infty)$ to $[0,1]$.
In so doing, we ensure that $\heps^+$ is bounded within $[0,0.5]$ and $\heps^-$ within $[0.5,1]$.
Similarly, we apply model-specific transformations to the remaining latent variables so that they are in the correct domains.

To estimate the ELBO, $\mathbbm{E}_q [\log p(y, \theta) - \log q_\lambda (\theta)]$, we need to be able to sample from $p(y, \theta)$ and from $q_\lambda (\theta)$.
First, we can compute $\log p(y, \theta)$ from the PGABM.
Second, the variational distribution $q_{\lambda}$ is an arbitrary parametric distribution we can sample from.
Thus, we can estimate the ELBO with a Monte Carlo routine.
In our case, the observed data $y$ correspond to $s$, i.e., the outcome of the interactions.

To maximize the ELBO, we are required to estimate its gradient with respect to $\theta$ at each iteration of the optimization.
The Gumbel-Softmax relaxation allows us to compute this gradient for the categorical variables, while for the continuous variables we can rely on the typical steps of AD.

In all our experiments we adopt SVI to maximize the ELBO.
After tuning the hyperparameters, we opted for the Adam optimizer, with learning rate \num{0.01}, running for \num{20000} epochs for \svinorm.
For \sviNF, we chose Block Neural Autoregressive flow~\citep{de2020block}, with 2 flows, in the NFs transformation.
In this case, we run Adam for \num{10000} epochs.
    
\spara{Markov Chain Monte Carlo.}
An alternative approach to estimate the latent variables of the PGM is sampling-based inference via an MCMC algorithm~\cite{bishop2006pattern}.
We use the No-U Turn Sampler (NUTS), an adaptive Hamiltonian Monte Carlo method that is known for its high efficiency and flexibility~\citep{hoffman2014no}.

We use \num{5000} simulations as burn-in and \num{5000} simulations for the estimate. 
Note that these simulations are run from the PGABM, not from the ABM, and they leverage on the conditional independencies of the PGM.

\spara{Approximate Bayesian Computation.}
While the previous methods are based on PGMs, Approximate Bayesian Computation (ABC) is a likelihood-free, simulation-based method~\citep{csillery2010approximate}, widely used in ABMs~\cite{carrella2021no, grazzini2017bayesian}.
ABC samples a set of parameters from a prior distribution, and runs the entire simulation of the opinion dynamics model for each sample.
It then considers summary statistics of the generated data-trace to see how `close' it is to the original data.
In this case, we use the number of positive and negative interactions per time step as summary statistics.
The distance between the simulated time series and the data summary is the $L^2$ distance. 
If the distance between the observed and simulated summary statistics is lower than an acceptance threshold, then the simulation is accepted; otherwise, it is rejected.
\corr{We launch \num{10000} complete simulations, and we set the median distance as acceptance threshold.}
Finally, it derives the posterior distribution of the target parameters by counting the number of acceptances and rejections from the different regions of the parameter space via Bayes' rule.

\spara{Experimental Settings.}
In each scenario, we run a grid of experiments where we sample \eps, with $\eps^+$ in $\{0.05,\ldots,0.45\}$ and $\eps^-$ in $\{0.55,\ldots,0.95\}$, and the model-specific latent variables from their domain.
In all models, we set $\mu^+ = \mu^- = 0.02$ and use 10 interactions per time step, \corr{unless otherwise specified}.
The number interactions per time step does not affect the results of variational inference methods, as the interactions are conditionally independent and are treated separately.
We vary the length of the simulation, $T$, and one other relevant parameter (specific to each model).
For each combination of parameters, we simulate the ABM and then estimate the target parameters with the four methods.
For practical reasons, we abort an experiment after \timeout.
Since the methods under scrutiny are all Bayesian, the parameter estimates are distributions.
For easier understanding of the results, we compare the average values of \num{200} samples from each posterior distribution.\footnote{The code to reproduce these results is open-source and available at\\\url{https://github.com/jaclenti/VI_opinion_dynamics}}

\subsection{BCM-S.}
\label{sec:BCM-S}
\spara{Model.}
First, we define an opinion dynamics model with structural rule (\ODMS).
Inspired by \citet{weisbuch2002meet}, we assign each agent $u$ a unique role $r_u$ between \emph{leader} ($r_u = L$) and \emph{follower} ($r_u = F$).
Leaders and followers adopt different behaviors.
An interaction $(u,v)$ follows a BCM-b with parameters \epF and \emF if $v$ is a follower and with parameters \epL and \emL if $v$ is a leader.
\corr{We assume that followers are more likely to change their opinions}, and thus $\epF\geq\epL$, $\emF\leq\emL$, $\mpF\geq\mpL$, and $\mmF\geq\mmL$.
For conciseness, we define $\eps = (\epF, \epL, \emF, \emL)$, $\mu = (\mpF, \mpL, \mmF, \mmL)$ and $r$ the vector of agents' roles. 
In summary, at each time step, the model evolves according to the rules described in \Cref{alg:bcm-lr}.
In this scenario, the goal is to estimate both \eps and $r$.

\begin{algorithm}[h!]
\small
\begin{algorithmic}[1]
\STATE{Sample two agents $(u,v)$ uniformly at random}
\STATE{Let $r_v \in \{F,L\}$ be the role of $v$}
\IF{$\lvert \Delta x^t_{uv} \rvert \leq \eps^+_{r_v}$}
	\STATE{$s \coloneqq 1$}
	\STATE{$x^{t+1}_v \coloneqq x^t_v + \mu^+_{r_v} \Delta x^t_{uv} $}
\ELSIF{$\lvert \Delta x^t_{uv} \rvert \geq \eps^-_{r_v}$}
	\STATE{$s \coloneqq -1$}
	\STATE{$x^{t+1}_v \coloneqq x^t_v - \mu^-_{r_v} \Delta x^t_{uv}$}
\ELSE
	\STATE{$s \coloneqq 0$}
	\STATE{$x^{t+1}_v \coloneqq x^t_v$}
\ENDIF
	\STATE{$x^{t+1}_v \coloneqq \max (0, \min ( x^{t+1}_v, 1) )$}
\end{algorithmic}
\caption{BCM-S: BCM-b with roles (leader/follower).}
\label{alg:bcm-lr}
\end{algorithm}

\spara{PGABM.}
The first step toward inference consists in the translation of the ABM into a proper PGABM.
We assume to know the update rates $\mu$ and to observe the interaction outcomes $s$, the identity of the pairs of interacting agents $e$, and the initial opinions $X_0$.
The observed outcome of the interactions $s$ depends on the latent variables, $r$ and $\eps$. 
Thus, we need to rewrite the rules of the ABM to express $s$ as a conditional probability.
Defining $s_j = (s^+_j, s^-_j)$ the outcome of the interaction $e_j = (u,v)$, we can write
\begin{align}
    & P(s^+_j = 1 \mid x_t, e_j, \eps, r) = \sigma(\rho \cdot( \eps^+_{r_v} - \lvert \Delta x^t_{uv} \rvert)) \label{eq:kappa_plus_S},\\
    & P(s^-_j = 1 \mid x_t, e_j, \eps, r) = \sigma(-\rho \cdot(\eps^-_{r_v} - \lvert \Delta x^t_{uv} \rvert)) \label{eq:kappa_minus_S}.
\end{align}

\spara{Variational Inference.}
Upon sampling from $q_\lambda$, we need to restrict the parameters in the appropriate domains since $q_\lambda$ has support in $\mathbbm{R}^{N+4}$.
We compute $\hat{\eps}$ and $\hat{r}$ from $\theta \sim q_\lambda$ by using a sigmoid transform.
\begin{multicols}{2}
\begin{squishlist}
    \item $\hepF = \sfrac{\sigma(\theta_1)}{2}$
    {\nolinenumbers
    \item $\hepL = \sfrac{\sigma(\theta_2)}{2}$}
    {\nolinenumbers
     \item $\hemF = \sfrac{\sigma(\theta_3)}{2} + \sfrac{1}{2}$}
     \item $\hemL = \sfrac{\sigma(\theta_4)}{2} + \sfrac{1}{2}$
\end{squishlist}
\end{multicols}
\noindent This step ensures that \hepF and \hepL are bounded within $[0,0.5]$ and \hemF and \hemL within $[0.5,1]$.

The procedure for the role vector $\mathbf{r}$ is slightly more complex as it represents categorical variables.
For each user $u = 1,\ldots,N$, we define $\hat{\phi}_u = \sigma(\theta_{u+4})$  which represents the posterior probability that $u = L$.
Instead of sampling the role with Bernoulli extractions of $\hat{\phi}$, we use the Gumbel-Softmax relaxation, with temperature 0.1.
Consequently, we define $\Tilde{r}_u \sim \text{Gumbel-Softmax}(\phi_u, 1-\phi_u)$.
Finally, we can express the bounded confidence interval for $u$ as the expectation over its role $\Tilde{\eps}_u = \Tilde{r}_u \cdot \hat{\eps}_L + (1-\Tilde{r}_u) \cdot \hat{\eps}_F$, in a way that is similar to a linear relaxation or a mixed membership.
Note that this expression simply represents a superposition of the beliefs of the model given the current evidence for the role of an agent, but does not require changing the model to define an intermediate `in-between' role.
This formulation allows determining $\Tilde{\eps}_u$ based on the sampled Gumbel-Softmax probabilities and builds an interpolation that is easier to handle.
After casting the PGABM into a differentiable shape, we maximize the ELBO via stochastic gradient ascent.

\begin{figure}[t]
\centering
\includegraphics[width=\linewidth]{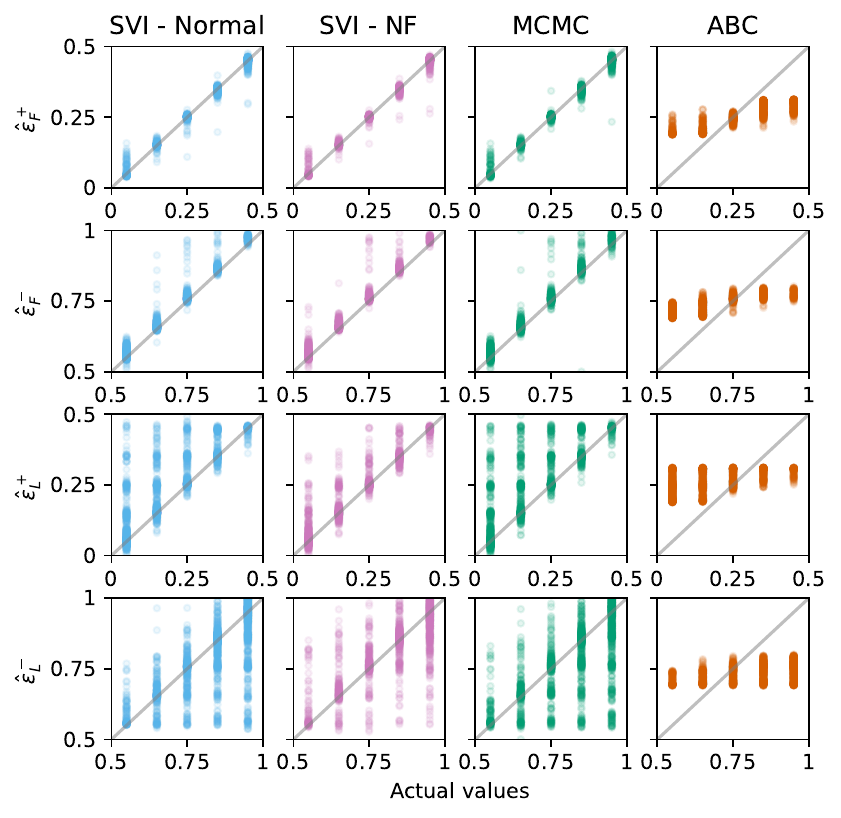}
\vspace{-\baselineskip}
\caption{
\ODMS.
Comparison between actual values \pmb{\eps} (x-axis) and estimates $\pmb{\hat{\eps}}$ (y-axis) for each macroscopic parameter (rows) and method (columns). Each experiment samples \pmb{\epF} and \pmb{\epL} in $\pmb{\{0.05, 0.15, 0.25, 0.35, 0.45\}}$, such that \pmb{\epF} $\pmb{\geq}$ \pmb{\epL}, \pmb{\emF} and \pmb{\emL} in $\pmb{\{0.55, 0.65, 0.75, 0.85, 0.95\}}$, and \pmb{\epF} $\pmb{\leq}$ \pmb{\epL}. 
Points on the diagonals represent exact estimates.}
\label{fig:BCMS_scatterplot_epsilon}
\end{figure}

\spara{Results.}
\Cref{fig:BCMS_scatterplot_epsilon} compares the estimates of the \emph{macroscopic} parameters with their actual values for all the experiments.
Summarizing the results, we obtain an average RMSE of \svinormAvgRMSE for \svinorm, of \sviNFAvgRMSE for \sviNF, \mcmcAvgRMSE for \mcmc and \abcAvgRMSE for \abc.
The improved performance of the SVI-based methods is evident even by inspecting the figure visually.

The main hurdle in the inference is discriminating between leaders and followers, both because of the high dimensionality of the parameter and its categorical nature.
Since most of the users are followers, the leader's parameters are the hardest to estimate: when the methods are not able to identify the leaders, they tend to set $\hepL = \hepF$ and $\hemL = \hemF$.
Since \epF $\geq$ \epL, several observations in \Cref{fig:BCMS_scatterplot_epsilon} (row 3) are above the diagonal.
Analogously, since $\emF \leq \emL$, most of the estimates in \Cref{fig:BCMS_scatterplot_epsilon} (row 4) are below the diagonal.

\begin{figure}[t]
\centering
\includegraphics[width=0.9\linewidth]{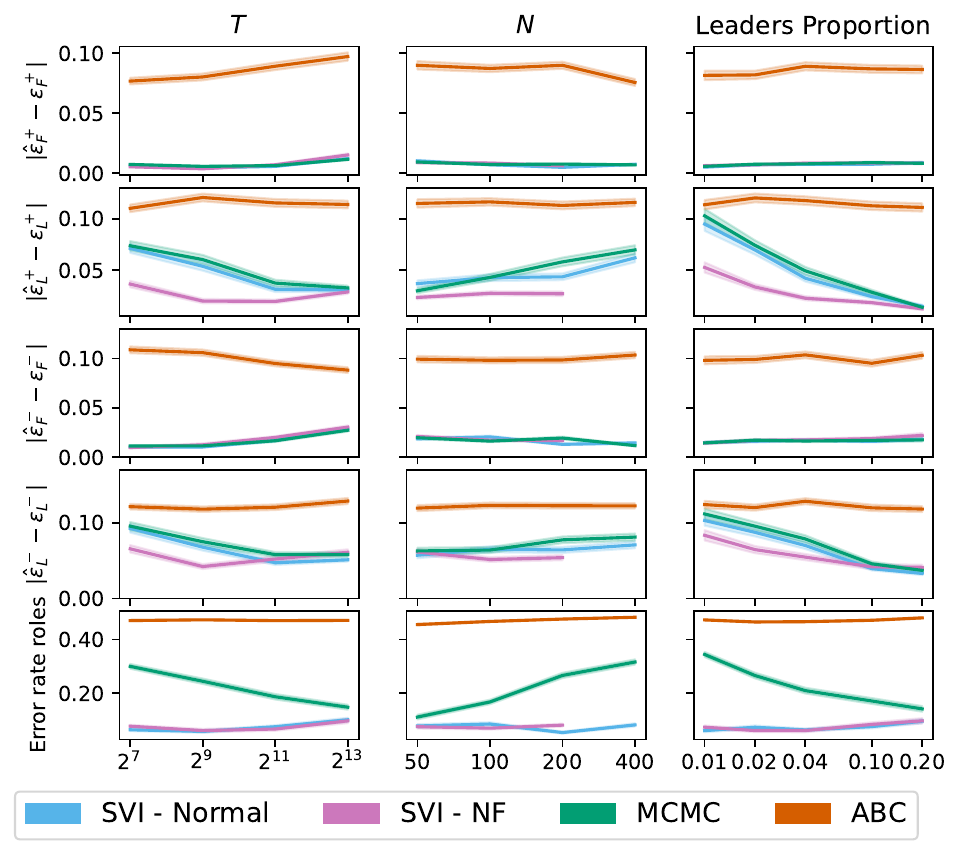}
\caption{
\ODMS. 
Specific errors on \pmb{\epF}, \pmb{\epL}, \pmb{\emF}, \pmb{\emL}, and \roles as functions of $\pmb{T}$ (left), $\pmb{N}$ (center), and proportion of leaders (right). The error bars represent the standard errors. 
}
\label{fig:BCMS_lineplot_performance}
\end{figure}

The error rate of the roles, and consequently all the other errors, grows as $N$ increases (\Cref{fig:BCMS_lineplot_performance}, column 1). 
Indeed, as $N$ grows, we have both an increased number of target parameters and a reduced number of observations per agent.
Conversely, as $T$ grows, we have more data, thus the estimates of SVI and MCMC improve (\Cref{fig:BCMS_lineplot_performance}, column 2).
A higher proportion of leaders leads to more accurate estimations (\Cref{fig:BCMS_lineplot_performance} column 3). 
Indeed, more leaders imply a larger number of interactions involving \epL and \emL, which helps in estimating the correct roles by differentiating the agents.
Moreover, the VI methods are the only ones able to estimate the role of microparameters with high accuracy (\Cref{fig:BCMS_lineplot_performance} row 4).
The average error rates for \svinorm, \sviNF, \mcmc, and \abc, are \svinormAvgErrorRoles, \sviNFAvgErrorRoles, \mcmcAvgErrorRoles, and \abcAvgErrorRoles, respectively.
The proportion of experiments with 100\% accuracy is, respectively, \svinormexactroles, \sviNFexactroles, \svimcmcexactroles, and \sviabcexactroles.

\subsection{BCM-I.} 
\label{sec:BCM-I}
\spara{Model.}
Second, we analyze an opinion dynamics model with interaction rule.
A relevant branch of opinion dynamics models consists of the models with synchronous interactions~\citep{hegselmann2002opinion}, where agents have many-to-one interactions instead of one-to-one interactions.
At each time step, an agent $v$ interacts with $K$ agents and gets influenced by all of them.
The dynamics of this variant of the BCM-b are based on the distance between $x_v$ and the average opinion of the other $K$ interacting agents.
We denote this distance with $\lvert \Delta x^{t}_{\overline{v} v} \rvert$.

An important question in these models with higher-order interactions is which order is the most influential for the opinion dynamics.
Indeed, given limited attention, only a subset of all the neighbors may have an effect~\citep{ceragioli2021modeling}.
We assume that at each interaction, $v$ and $l_v = (u_1,\ldots,u_F)$ are sampled, where $l_F$ is an ordered list of $F > K$ agents (e.g., the agents that compose a social media feed).
Then, the dynamics of the BCM-b are followed based on $x_v$ and the average opinion of the first $K$ elements of $l_F$ (i.e., the subset that $v$ pays attention to), as illustrated in \Cref{alg:bcm-f}.
This model simulates a realistic social media scenario in which a user is potentially exposed to long feeds of content but is influenced only by a fraction of it.
In this model, our goal is to estimate \eps and $K$.

\begin{algorithm}[h!]
\small
\begin{algorithmic}[1]
\STATE{Sample F+1 agents $(u_1, \ldots,u_F, v)$ uniformly at random}
\STATE{Let $\Delta x^{t}_{\overline{v} v} = x^t_v - \frac{1}{K}\sum_{j\leq K}x^t_{v_j}$}
\IF{$\lvert \Delta x^{t}_{\overline{v} v} \rvert \leq \eps^+$}
	\STATE{$s \coloneqq 1$}
	\STATE{$x^{t+1}_v \coloneqq x^t_v + \mu^+ \Delta x^{t}_{\overline{v} v} $}
\ELSIF{$\lvert \Delta x^{t}_{\overline{v} v} \rvert \geq \eps^-$}
	\STATE{$s \coloneqq -1$}
	\STATE{$x^{t+1}_v \coloneqq x^t_v - \mu^- \Delta x^{t}_{\overline{v} v}$}
\ELSE
	\STATE{$s \coloneqq 0$}
	\STATE{$x^{t+1}_v \coloneqq x^t_v$}
\ENDIF
	\STATE{$x^{t+1}_v \coloneqq \max (0, \min ( x^{t+1}_v, 1) )$}
\end{algorithmic}
\caption{BCM-I: BCM-b with higher order interactions.}
\label{alg:bcm-f}
\end{algorithm}

\spara{PGABM.}
In this scenario, the parameter $e$ comprises all the potential interacting agents, $e_j = (u_1, \ldots, u_F, v)$.
Thus, we need to include the observed parameter $F$, the latent variable $K$ and the variable $\tilde{e}$, which is the result of a deterministic function of $e$ and $K$, representing the actual agents that interact.

The conditional probabilities of this model are
\begin{align}
    & P(s^+_j = 1 \mid x_t, e_j, \eps, K) = \sigma(\rho \cdot( \eps^+ - \lvert \Delta x^{t}_{\overline{v} v} (K) \rvert)) \label{eq:kappa_plus_I}\\
    & P(s^-_j = 1 \mid x_t, e_j, \eps, K) = \sigma(-\rho \cdot(\eps^-_{r_v} - \lvert \Delta x^{t}_{\overline{v} v} (K) \rvert)), \label{eq:kappa_minus_I}
\end{align}
where we write the distance between the opinion of $v$ and the interacting agents as a function of $K$ ($\lvert \Delta x^{t}_{\overline{v} v} (k) \rvert$), thus highlighting the dependence on $K$ of the opinion average.

\spara{Variational Inference.}
In this second scenario, the introduction of the latent categorical variable $K$ hinders the differentiability of the model.
$\theta \sim q_{\lambda}$ has dimension $2 + F$.
The first two elements of each sample, $\theta_1$ and $\theta_2$, are associated to $\eps^+$ and $\eps^-$, while the remaining $F$, $\theta_3, \ldots, \theta_F$, are associated to the possible values of $K$ (represented as a categorical).
$\hat{\eps}^+$ and $\hat{\eps}^-$ are computed from $\theta_1$ and $\theta_2$ as in BCM-b.
For $j > 2$, we bound $\theta_j$ within $[0,1]$, through $\phi_j = \sigma(\theta_{j + 2})$.
In this sense, $\hat{\phi}_j$ is our estimated posterior probability that $K$ is equal to $j$.
Instead of sampling $K \sim \text{Categorical}(\phi_1,\ldots,\phi_F)$, we write $K \sim \text{Gumbel-Softmax}(\phi_1, \ldots, \phi_F)$.

Now, let $\kappa(\eps, K)$ be the probability of observing $s$ as a function of $\eps$ and $K$ (\Cref{eq:kappa_plus_I} and \Cref{eq:kappa_minus_I}).
We define $\tilde{\kappa}(\eps) = \sum_j \phi_j \kappa(\eps, j)$ and compute the $p(y, \theta)$ of the ELBO (\Cref{eq:ELBO}) from $\tilde{\kappa}(\eps)$.
Hence, the ELBO aims to maximize the expected value on $K$ of the probability of observing $s$.

\begin{figure}[t]
\centering
\includegraphics[width=0.7\linewidth]{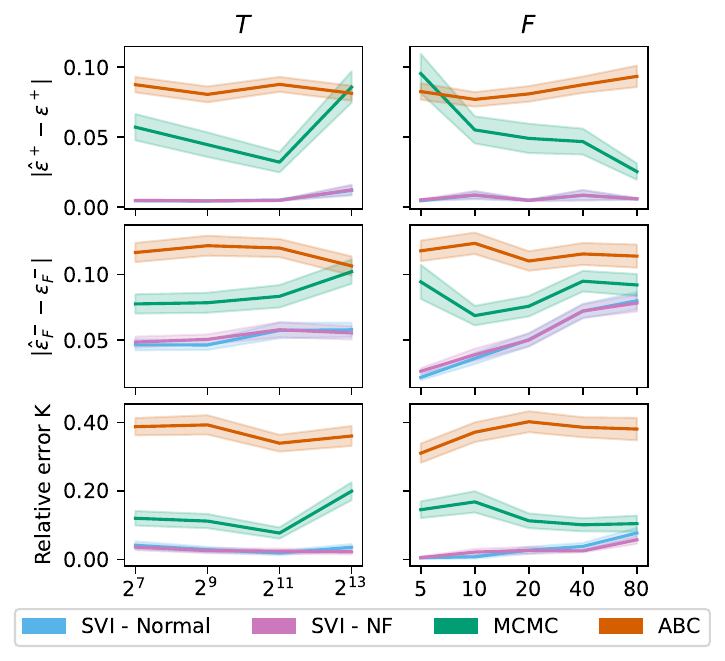}
\caption{
\ODMI.
Specific errors on $\pmb{\eps^+}$, $\pmb{\eps^-}$, and $\pmb{K}$, as functions of $\pmb{T}$ (left), $\pmb{F}$ (right). 
The error bars represent the standard errors. 
The relative error on $K$ is computed as $\lvert \hat{K} - K \rvert / F$.}
\label{fig:BCMI_lineplot_performance}
\end{figure}

\spara{Results.}
We vary $T \in \{128, 512, 2048, 8192\}$, and $F \in \{5, 10, 20, 40, 80\}$, while we keep $N=400$.
For the estimates of \eps, VI with normal distribution shows the best performances, improving the RMSE of VI with NFs by a factor of \IRMSEsviNFvssvinormal, MCMC by \IRMSEmcmcvssvinormal, and ABC by \IRMSEabcvssvinormal (\Cref{fig:BCMI_lineplot_performance}, rows 1 and 2).
Overall, considering only the experiments where all methods terminate successfully, the relative errors on $K$ are respectively, \IerrFsvinormal, \IerrFsviNF, \IerrFmcmc and \IerrFabc (\Cref{fig:BCMI_lineplot_performance}, rows 3).
VI is one order of magnitude more accurate than ABC for this model.

We can observe increasing errors on $\eps^-$ as $F$ increases.
Indeed, increasing $F$ (and, consequently, $K$) dampens the effect of the extreme opinions by taking into account a larger set of agents in the computation of the mean opinion.
This in turn results in a reduced number of negative interactions, posing a problem for the practical identifiability of $\eps^-$.

\subsection{BCM-U.}
\label{sec:BCM-U}
\spara{Model.}
Third, we focus on an opinion dynamics model with update rule (\ODMU).
The BCM-b is comprised of two possible updates: the convergence and the divergence of the agents' opinions, as a consequence of a positive or negative interaction.

We consider an extended version that depends on a global binary variable $\beta$. 
If $\beta = 1$, the agents update their opinion with backfire effect, otherwise if $\beta = 0$, their opinion remains unchanged after a negative interaction.
In this context, we aim to learn \eps and $\beta$.
This means that we ask whether the backfire effect is present, by observing the interactions, \corr{the signs of their outcomes}, and the initial opinions.
The model evolves as shown by \Cref{alg:bcm-b}.

\begin{algorithm}[h!]
\small
\begin{algorithmic}[1]
\STATE{Sample two agents $(u,v)$ uniformly at random}

\IF{$\lvert \Delta x^t_{uv} \rvert \leq \eps^+$}
	\STATE{$s \coloneqq 1$}
	\STATE{$x^{t+1}_v \coloneqq x^t_v + \mu^+_{r_v} \Delta x^t_{uv} $}
\ELSIF{$\lvert \Delta x^t_{uv} \rvert \geq \eps^-$}
	\STATE{$s \coloneqq -1$}
	\STATE{$x^{t+1}_v \coloneqq x^t_v - \beta \mu^-_{r_v} \Delta x^t_{uv}$}
\ELSE
	\STATE{$s \coloneqq 0$}
	\STATE{$x^{t+1}_v \coloneqq x^t_v$}
\ENDIF
	\STATE{$x^{t+1}_v \coloneqq \max (0, \min ( x^{t+1}_v, 1) )$}
\end{algorithmic}
\caption{BCM-U: BCM-b with modulated backfire.}
\label{alg:bcm-b}
\end{algorithm}

\spara{PGABM.}
In this case we extend the PGM of the BCM-b, by including the latent variable $\beta$ that affects $X_t$.
The conditional probabilities of the PGABM are similar to the ones in BCM-b (\Cref{eq:kappa_plus_BCMb} and \Cref{eq:kappa_minus_BCMb}), 
\begin{align}
    & P(s^+_j = 1 \mid x_t, e_j, \eps, \beta) = \sigma(\rho \cdot( \eps^+ - \lvert \Delta x^t_{uv}(\beta) \rvert)) \label{eq:kappa_plus_U},\\
    & P(s^-_j = 1 \mid x_t, e_j, \eps, \beta) = \sigma(-\rho \cdot(\eps^- - \lvert \Delta x^t_{uv}(\beta) \rvert)) \label{eq:kappa_minus_U},
\end{align}
but $x^t_{uv}(\beta)$ depends on $\beta$.

\spara{Variational Inference.}
The variational distribution $q_\lambda$ has dimension 3.
Given $\theta \sim q_\lambda$, $\theta_1$ and $\theta_2$ refer to \eps, in the same manner as previous models, while $\theta_3$ encodes the knowledge on $\beta$.
We aim to learn the posterior distribution of the binary variable $\beta$.
Therefore, $\phi = \sigma(\theta_3)$ to have a plausible probability.
Similarly to previous scenarios, we make the model differentiable via reparameterization, $\hat{\beta} \sim \text{Gumbel-Softmax}(\phi, 1 - \phi)$.
Then, we denote with $\kappa(\eps, \beta)$ the probability of observing $s$ as a function of $\beta$.
In analogy with \Cref{sec:BCM-I}, we inject the ELBO with $\tilde{\kappa}(\eps) = \phi \kappa(\eps, 0) + (1 - \phi) \kappa(\eps,1)$.
In this case, this operationalization is feasible.
The opinions $X_t$ depend deterministically on $\beta$ (which can be 0 or 1), on $s$ and $\mu$, which are observed, and on $X_{t-1}$. 
Thus, we can have only two possible realizations of $X_t$, for $\beta$ equal to 0 and 1.
So, we compute beforehand $x_t^{\beta = 0}$ and $x_t^{\beta = 1}$, and consequently $\Delta x^t_{uv}(\beta)$.
Conversely, different update rules would lead to more complex evolutions of $X_t$, hindering the ability to compute in advance all the possible realizations of $X_t$.

\begin{figure}[t]
\centering
\includegraphics[width=0.9\linewidth]{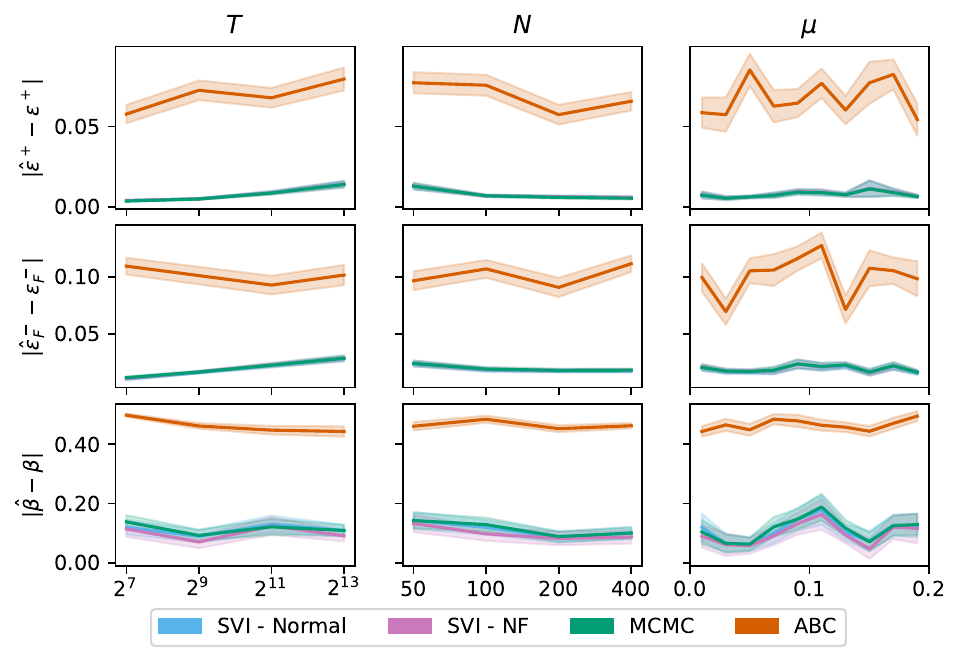}

\caption{\ODMU.
Specific errors on $\pmb{\eps^+}$, $\pmb{\eps^-}$, and $\pmb{\beta}$, as functions of $\pmb{T}$ (left), $\pmb{N}$ (center) and $\pmb{\mu}$ (right).
The error bars represent the standard errors.
}
\label{fig:BCMU_lineplot_performance}
\end{figure}

\spara{Results.}
We run a grid of experiments varying $T \in \{128, 512, 2048, 8192\}$, $N \in \{50, 100, 200, 400\}$, and $\mu^+ = \mu^- = \{0.01, 0.03,\ldots,0.19\}$.
The experiments in this scenario show higher accuracies than the previous ones, and the performance of the three PGM-based methods is similar (\Cref{fig:BCMU_lineplot_performance}).
Instead, ABC provides a poor estimate of $\beta$ in all settings.
The average error on $\beta$ of the four models is, respectively, \Ubetasvinormal, \UbetasviNF, \Ubetamcmc, and \Ubetaabc.
Similarly, VI with Normal improves the RMSE(\eps, $\hat{\eps}$) by a factor of \URMSEsviNFvssvinormal compared to VI with NFs, \URMSEmcmcvssvinormal to MCMC and \URMSEabcvssvinormal to ABC.
These results reflect the similar performances of VI and MCMC in this simpler scenario and the lower performance of ABC.

\subsection{BCM-G.}
\label{sec:BCM-G}
\spara{Model.}
Finally, we define an opinion dynamics model with a graph rule.
Taking inspiration from previous literature~\citep{de2022modelling}, we consider the possibility of link rewiring after a negative interaction.
In this extension, an interaction $(u,v)$ can have two dynamics.
With probability \corr{$\xi$} the agents follow the dynamics of opinion update, as in BCM-b.
With probability \corr{$1 - \xi$}, they follow the dynamics of link rewiring, meaning that if \corr{$\lvert \Delta x^t_{uv}\rvert > \gamma$} the link $(u,v)$ is broken.
In this latter case, the link $(u,v)$ and a random link $(w,z)$ are severed and the links $(u,z)$ and $(w,v)$ are created.
This way, the edge density and the degree sequence are preserved by each rewiring.
The sampled nodes $(w,z)$ are selected to maintain the connectedness of the graph.
To allow for the possibility of creating and removing links, we need to define the social network in advance.
Thus, at the beginning of the simulations we connect the agents with an Erd\H{o}s-R\'{e}nyi with density 0.1, as shown in \Cref{alg:bcm-rewiring}.
In this model the goal is to estimate \eps and $\gamma$.

\begin{algorithm}[h!]
\small
\begin{algorithmic}[1]
\STATE{Sample two connected agents $(u,v)$ uniformly at random}
\STATE{Sample $h \in [0,1]$}
\IF{$h <$\corr{$ \xi$}}
    \IF{$\lvert \Delta x^t_{uv} \rvert \leq \eps^+$}
    	\STATE{$s \coloneqq 1$}
    	\STATE{$x^{t+1}_v \coloneqq x^t_v + \mu^+_{r_v} \Delta x^t_{uv} $}
    \ELSIF{$\lvert \Delta x^t_{uv} \rvert \geq \eps^-$}
    	\STATE{$s \coloneqq -1$}
    	\STATE{$x^{t+1}_v \coloneqq x^t_v - \beta \mu^-_{r_v} \Delta x^t_{uv}$}
    \ELSE
    	\STATE{$s \coloneqq 0$}
    	\STATE{$x^{t+1}_v \coloneqq x^t_v$}
    \ENDIF
\ELSE
    \IF{$\lvert \Delta x^t_{uv} \rvert > \gamma$}
    	\STATE{remove $\overline{uv}$}
        \STATE{Sample two connected agents $(w,z)$ uniformly at random}
    	\STATE{remove $\overline{wz}$}
        \STATE{add $\overline{uz}$}
    	\STATE{add $\overline{wv}$}
    \ENDIF
\ENDIF    
\STATE{$x^{t+1}_v \coloneqq \max (0, \min ( x^{t+1}_v, 1) )$}
\end{algorithmic}
\caption{BCM-G: BCM-b with link rewiring.}
\label{alg:bcm-rewiring}
\end{algorithm}

\spara{PGABM.}
In this scenario, we assume to observe all the interactions and their outcomes, both for the interactions with opinion update and the ones with rewiring dynamics.
Moreover, we assume to know \corr{$\xi$}.
We define a variable $d$ that determines the dynamics of the interaction: it is 0 if the interaction has update dynamics and 1 if it has rewiring dynamics.
So, an interaction between $u$ and $v$ has outcome $s = (s^+, s^-, s^r, d)$.
If $d = 0$, $s^+$ and $s^-$ encode the sign of the update interaction, while if $d = 1$, $s^r$ encodes the outcome of the rewiring interaction.

In this scenario, we have the same conditional probabilities of the BCM-b (\Cref{eq:kappa_plus_BCMb} and \Cref{eq:kappa_minus_BCMb}), plus the dynamics of rewiring, that have the same form:  
\begin{align}
    & P(s^+_j = 1 \mid x_t, e_j, \eps, \gamma) = \sigma(\rho \cdot( \eps^+ - \lvert \Delta x^t_{uv} \rvert)) (1 - d_j)\label{eq:kappa_plus_G},\\
    & P(s^-_j = 1 \mid x_t, e_j, \eps, \gamma) = \sigma(-\rho \cdot(\eps^- - \lvert \Delta x^t_{uv} \rvert)) (1 - d_j)\label{eq:kappa_minus_G},\\
    & P(s^r_j = 1 \mid x_t, e_j, \eps, \gamma) = \sigma(-\rho \cdot(\gamma - \lvert \Delta x^t_{uv} \rvert)) d_j\label{eq:kappa_r_G},
\end{align}

\spara{Variational Inference.}
There are no discrete variables in the model, so we do not need additional approximations.
The variational distribution $q_\lambda$ has dimension 3.
The parameters $\theta_1$ and $\theta_2$ that are associated to \eps have the usual bounds, while $\theta_3$ is bounded within $[0,1]$.
Hence, $\hat{\gamma} = \sigma(\theta_3)$.
Having only continuous variables, the estimation of the ELBO and its gradient is straightforward.

\spara{Results.}
In the experiments, we vary $T \in \{128, 512, 2048, 8192\}$ and \corr{$\xi$}$ \in \{0.2, 0.4, 0.6, 0.8\}$.
Again, the PGABM-based techniques overperform ABC, with an average RMSE of \GRMSEsvinormal, \GRMSEsviNF, \GRMSEmcmc, \GRMSEabc for the four methods.
In particular, the accuracy of $\hat{\gamma}$ depends on the number of observations of rewiring, improving with higher $T$ (more interactions) and lower \corr{$\xi$} (higher proportion of rewiring interactions).

 \begin{figure}[t]
\centering
\includegraphics[width=0.7\linewidth]{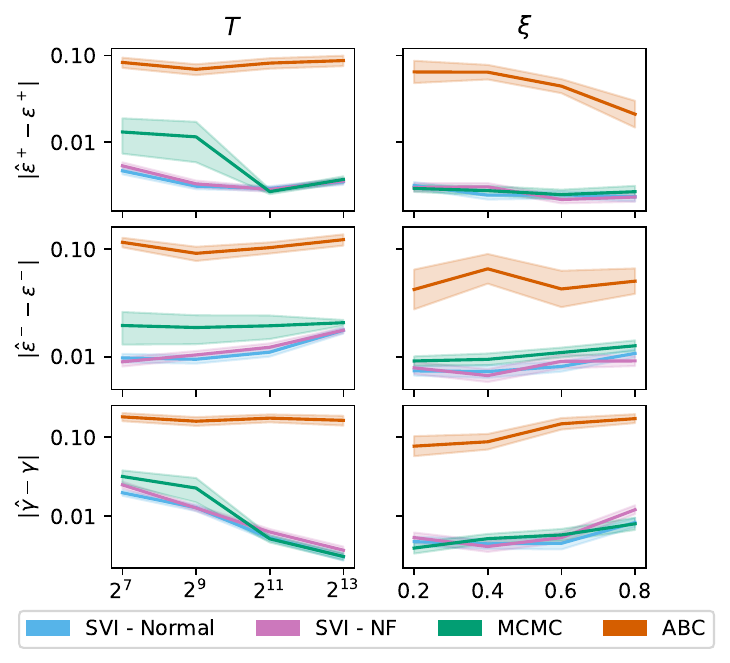}
\caption{
\ODMG.
Specific errors on $\pmb{\eps^+}$, $\pmb{\eps^-}$, and $\pmb{\gamma}$,
as functions of $\pmb{T}$ (left), \corr{$\pmb{\xi}$} (right). 
The error bars represent the standard errors. 
}
\label{fig:BCMG_lineplot_performance}
\end{figure}

\subsection{Running time.}

 \begin{figure*}[t]
\centering
\includegraphics[width=0.85\linewidth]{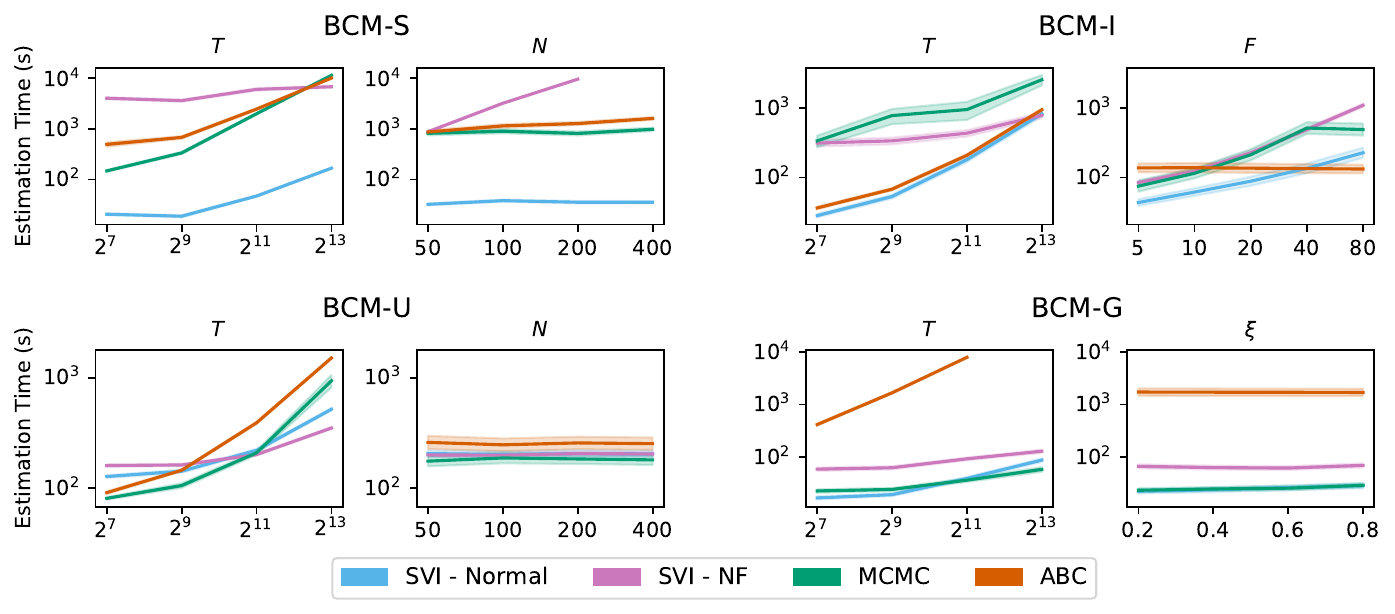}
\caption{
Estimation time measured as a function of the main hyperparameters of our experiments.
In PGM-based methods, estimation time increases with the number of latent variables, $N$ in BCM-S and $F$ in BCM-I. 
In ABC it, depends on the length of the simulations, $T$.
}
\label{fig:BCMS_lineplot_time}
\end{figure*}

The estimation times vary considerably among the tested methods (\Cref{fig:BCMS_lineplot_time}).
First, both the VI methods scale well with the length of the ABMs trajectories ($T$).
The scalability of VI arises from its optimization-based approach, which avoids the need to explicitly evaluate the entire dataset during inference.
This computational advantage becomes more evident as the dataset size grows~\citep{blei2017variational}.
Second, the estimation time of \sviNF grows linearly with the number of parameters.
This reflects the increased number of parameters underlying the neural networks that implements the NFs.
This is evident in \ODMS, where the number of parameters is equal to $N + 4$.
Conversely, the estimation time of MCMC  and ABC increases linearly with $T$, reflecting the number of samples required to replicate larger datasets.
Overall, \svinorm is the most efficient method: in \ODMS, it reduces the estimation time by a factor of \svinormvssviNFtime compared to \sviNF, \svinormvsMCMCtime compared to \mcmc, and \svinormvsABCtime compared to \abc.
Note that the experiments exceeding the time limit were aborted, but \svinorm was the only method that never exceeded the time limit.


\section{Discussion}
Our work marks a significant advancement in the design of data-driven realistic models for human behavior.
A computational approach to understand social systems passes through interpretable models, able to explain the causal effects leading to observed phenomena.
Moreover, employing causal models enables the analysis of what-if scenarios.
Such explanatory models are needed by policy-makers to evaluate interventions to address pressing social challenges.

Thanks to their mechanistic nature, Agent-Based Models (ABMs) depict a clear causal connection between individual actions and global phenomena.
However, the lack of connection between ABMs and data has so-far limited the use of these models in real-world scenarios.
Existing approaches to use available data in ABMs aim to roughly replicate some macroscopic properties, with poor statistical robustness.
This gap between ABMs and data can be related to several reasons.

\spara{ABMs simulations are computationally expensive.}
Our technique significantly reduces the computational burden, by tackling the inference problem as an optimization one.
Instead of massively running the ABMs to explore the properties arising from the simulations, we exploit the probabilistic nature of PGABMs to formulate a Variational Inference problem.
Minimal approximations of the ABM rules allow the model to be differentiable and leverage Automatic Differentiation (AD) tools.
These techniques drastically reduce the computational time when dealing with longer data trajectories.

Our approach inverts the problem by seeking the parameters that generated the observed data with the highest probability.
This paradigm also enables learning high-dimensional parameters.
We prove the ability to recover up to $400$ microscopic (agent-level) parameters.
This task is unfeasible with any simulation-based approach, as it requires exploring $2^{400}$ combinations of parameters and running a set of simulations for each combination.

\spara{ABMs lack a unique formalism.}
A shared formalism for describing ABMs is missing, leading to a deluge of ad-hoc solutions to calibrate and analyze them~\citep{daly2022quo}.
In our work, we adopt the PGABM representation.
The generative nature of ABMs makes the translation to PGABMs natural and opens the way to the wide literature on probabilistic models.
We remark that this general formalism can be employed in any ABM.

\spara{Equifinality.}
Traditional simulation-based methodologies use data summarization into low-dimensional statistics.
Although this dimensionality reduction eases comparing between observed and simulated traces, it can result in equifinality problems~\cite{an2023modeling}, where similar summary statistics can arise from different microscopic trajectories~\cite{anscombe1973graphs}.\footnote{In ML literature, this phenomenon is referred to as identifiability.}
Our approach avoids the loss of information induced by summarization by using the whole data trajectories when computing the ELBO.

\spara{Multifinality.}
The inherent stochasticity of the ABMs can lead to multifinality issues, as the same input parameters can produce diverse macroscopic properties of the system~\cite{an2023modeling}.
This feature is problematic in simulation-based calibration.
Indeed, one could obtain different results by calibrating the parameters with different initialization seeds.
Our approach alleviates this issue, by seeking the set of parameters that most likely generated the observed trace, instead of trying to replicate similar statistics.
This way, we take into account the stochasticity of the model, without the need to obtain the summary statistics of the data.






\spara{\emph{Limitations}.}
This work is not intended to be a universal solution for calibrating any ABM, but an extensive proof-of-concept of a novel and effective inference methodology.

The huge literature on ABMs has proposed an incredible variety of rules, that are often very complex and non-tractable, and we do not guarantee that the combination of VI and PGABMs could be properly adapted to any possible ABM.
However, the current work shows that this technique can be used in a wide range of rules for opinion dynamics models, targeting discrete, continuous, and high-dimensional parameters.
We encourage further research in the field to understand the limits and the potentials of PGABMs.
Some discrete distributions add layers of complexity that could likely become intractable, such as stochastic rules for ranking or matching.
However, most of the usual rules adopted in opinion dynamics literature are similar to the ones studied here.

All the models presented in the paper have longitudinal observations over time.
This creates conditional independencies that are efficiently exploited in the inference. 
This assumption is not necessary for applying Variational Inference, but it alleviates the computational burden, as it is not necessary to reproduce the complete trajectories during the training, but only the updates at a single time step.
Hence, the capability to infer the parameters from data trajectories depends on data availability.
While it can be straightforward when longitudinal microscopic data are available, it can be harder in the presence of longitudinal latent variables.
Fortunately, microscopic temporal data are becoming more common in many real social studies.
Regrettably, traditional calibration approaches employ summary statistics to reduce the dimensionality of the dataset, thus discarding more granular but valuable information.

Finally, this work focuses on synthetic data generated from ABMs.
The applicability of these techniques on real-world data needs to be understood.
However, we showed that they correctly recover the parameters in controlled experiments where the ground truth is known, which is the only systematic way to evaluate inference methods.
We do not assume that the presented opinion dynamics models realistically describe real-world interactions.
But, if we assume that the ABM can be used to describe the data generating process, then our approach can estimate the most likely parameters of the model.






\bibliographystyle{IEEEtranNAT}
\bibliography{bibliography}

\end{document}

%% file: macros.tex
\usepackage[switch]{lineno}

\usepackage[utf8]{inputenc} 
\usepackage[T1]{fontenc} 
\usepackage{amsmath,amsfonts}
\usepackage{array}
\usepackage[caption=false,font=normalsize,labelfont=sf,textfont=sf]{subfig}
\usepackage{textcomp}
\usepackage{stfloats}
\usepackage{url}
\usepackage{verbatim}
\usepackage{cite}

\usepackage{systeme}
\usepackage{bbm}
\usepackage{tikz}
\usepackage{wrapfig}
\usetikzlibrary{bayesnet}
\usepackage{dsfont}
\usepackage{comment}
\usepackage{type1cm} 
\usepackage{graphicx} 
\usepackage{xspace} 
\usepackage{balance} 
\usepackage{booktabs} 
\usepackage{multirow} 

\usepackage{bold-extra} 
\usepackage{microtype} 
\usepackage{siunitx} 
\usepackage{xfrac} 
\usepackage{mathtools} 
\usepackage{xspace} 
\PassOptionsToPackage{hyphens}{url} 
\usepackage[bookmarks, pdftex, colorlinks=true, pagebackref=true, backref=false]{hyperref} 
\usepackage{cleveref} 
\usepackage[square,sort,comma,numbers]{natbib} 
\usepackage{hyphenat} 
\usepackage[show]{chato-notes}
\usepackage[colaction]{multicol}
\usepackage{algorithm}
\usepackage{algorithmic}

\usepackage{placeins}
\tikzstyle{detobs} = [det, fill = gray!30]

\graphicspath{{images}}
\usepackage{adjustbox}

\newcommand{\spara}[1]{\smallskip\noindent\textbf{#1}}

\newenvironment{squishlist}
{\begin{list}{$\bullet$}
 {\setlength{\itemsep}{0pt}
     \setlength{\parsep}{3pt}
     \setlength{\topsep}{3pt}
     \setlength{\partopsep}{0pt}
     \setlength{\leftmargin}{1.5em}
     \setlength{\labelwidth}{1em}
     \setlength{\labelsep}{0.5em} } }
{\end{list}}

\usepackage{xcolor}
\hypersetup{
bookmarks, pdftex,
colorlinks=true,
pagebackref=true, backref=page,
linkcolor={red!50!black},
filecolor={green!50!black},
citecolor={green!50!black},
urlcolor={blue!80!black},
}

\newcommand{\ODMS}{BCM-S\xspace}
\newcommand{\ODMI}{BCM-I\xspace}
\newcommand{\ODMU}{BCM-U\xspace}
\newcommand{\ODMG}{BCM-G\xspace}

\newcommand{\epL}{\ensuremath{\varepsilon^+_L}\xspace}
\newcommand{\epF}{\ensuremath{\varepsilon^+_F}\xspace}
\newcommand{\emL}{\ensuremath{\varepsilon^-_L}\xspace}
\newcommand{\emF}{\ensuremath{\varepsilon^-_F}\xspace}
\newcommand{\heps}{\ensuremath{\hat{\varepsilon}}\xspace}
\newcommand{\hepL}{\ensuremath{\hat{\varepsilon}^+_L}\xspace}
\newcommand{\hepF}{\ensuremath{\hat{\varepsilon}^+_F}\xspace}
\newcommand{\hemL}{\ensuremath{\hat{\varepsilon}^-_L}\xspace}
\newcommand{\hemF}{\ensuremath{\hat{\varepsilon}^-_F}\xspace}
\newcommand{\mpL}{\ensuremath{\mu^+_L}\xspace}
\newcommand{\mpF}{\ensuremath{\mu^+_F}\xspace}
\newcommand{\mmL}{\ensuremath{\mu^-_L}\xspace}
\newcommand{\mmF}{\ensuremath{\mu^-_F}\xspace}
\newcommand{\bfeps}{\ensuremath{\pmb{\varepsilon}}\xspace}
\newcommand{\eps}{\ensuremath{\varepsilon}\xspace}
\newcommand{\roles}{\ensuremath{\mathbf{r}}\xspace}

\newcommand{\svinorm}{SVI with normal\xspace}
\newcommand{\sviNF}{SVI with NFs\xspace}
\newcommand{\mcmc}{MCMC\xspace}
\newcommand{\abc}{ABC\xspace}

\newcommand{\svinormvssviNFtime}{\ensuremath{56.0}\xspace}
\newcommand{\svinormvsMCMCtime}{\ensuremath{36.9}\xspace}
\newcommand{\svinormvsABCtime}{\ensuremath{36.5}\xspace}

\newcommand{\svinormAvgErrorRoles}{\ensuremath{0.09}\xspace}
\newcommand{\sviNFAvgErrorRoles}{\ensuremath{0.060}\xspace}
\newcommand{\mcmcAvgErrorRoles}{\ensuremath{0.22}\xspace}
\newcommand{\abcAvgErrorRoles}{\ensuremath{0.47}\xspace}
\newcommand{\svinormAvgRMSE}{\ensuremath{0.044}\xspace}
\newcommand{\sviNFAvgRMSE}{\ensuremath{0.036}\xspace}
\newcommand{\mcmcAvgRMSE}{\ensuremath{0.051}\xspace}
\newcommand{\abcAvgRMSE}{\ensuremath{0.125}\xspace}
\newcommand{\timeout}{\ensuremath{3} hours\xspace}
\newcommand{\svinormexactroles}{\ensuremath{0.38}\xspace}
\newcommand{\sviNFexactroles}{\ensuremath{0.54}\xspace}
\newcommand{\svimcmcexactroles}{\ensuremath{0.29}\xspace}
\newcommand{\sviabcexactroles}{\ensuremath{0}\xspace}

\newcommand{\IerrFsvinormal}{\ensuremath{0.031}\xspace}
\newcommand{\IerrFsviNF}{\ensuremath{0.027}\xspace}
\newcommand{\IerrFmcmc}{\ensuremath{0.126}\xspace}
\newcommand{\IerrFabc}{\ensuremath{0.370}\xspace}

\newcommand{\IRMSEsviNFvssvinormal}{\ensuremath{1.02}\xspace}
\newcommand{\IRMSEmcmcvssvinormal}{\ensuremath{2.13}\xspace}
\newcommand{\IRMSEabcvssvinormal}{\ensuremath{2.91}\xspace}

\newcommand{\Ubetasvinormal}{\ensuremath{0.112}\xspace}
\newcommand{\UbetasviNF}{\ensuremath{0.100}\xspace}
\newcommand{\Ubetamcmc}{\ensuremath{0.115}\xspace}
\newcommand{\Ubetaabc}{\ensuremath{0.465}\xspace}

\newcommand{\URMSEsviNFvssvinormal}{\ensuremath{1.02}\xspace}
\newcommand{\URMSEmcmcvssvinormal}{\ensuremath{1.02}\xspace}
\newcommand{\URMSEabcvssvinormal}{\ensuremath{6.11}\xspace}

\newcommand{\GRMSEsvinormal}{\ensuremath{0.009}\xspace}
\newcommand{\GRMSEsviNF}{\ensuremath{0.009}\xspace}
\newcommand{\GRMSEmcmc}{\ensuremath{0.016}\xspace}
\newcommand{\GRMSEabc}{\ensuremath{0.102}\xspace}

\newcommand{\corr}[1]{#1}

%% file: pgms_BCM.tex
\begin{figure}[t]
\vspace{-\baselineskip}
\begin{center}
      \begin{tikzpicture}
      \scalebox{0.8}{
        \node[det]                          (X)     {$\mathbf{X}_t$};
        \node[obs, left=0.4cm of X]     (X0)    {$\mathbf{X}_0$};
        \node[det, right=3cm of X]        (Xt1)   {$\mathbf{X}_{t+1}$};
        \node[obs, right=1.2cm of X, yshift = 0.7cm]        (s)     {$s$};
        \node[const, left=1cm of s, yshift=0cm]     (e)     {$e$};
        \node[latent, right=0.8cm of e, yshift=1.2cm]     (eps)     {\eps};
        \node[const, right=0.7cm of eps, yshift=0.cm]     (mu)     {$\mu$};

        \plate [minimum size=1.5cm, minimum width=4.2cm] {} {(s)
        (e) (X) (Xt1) } {$T$} ;

        \edge {e} {s}
        \edge {X} {s}
        \edge {X0} {X}
        \edge {X} {Xt1}
        \edge {s} {Xt1}
        \edge {mu} {Xt1}
        \edge {eps} {s}
      }
      \end{tikzpicture}
      \label{fig:pgm_bcmb}

\caption{Probabilistic Graphical Model associated with the Bounded Confidence model with backfire effect. Circles represent stochastic variables, diamonds deterministic variables, and letters without enclosures are given PGM parameters. Shaded variables are observed and white ones are latent. $\pmb{X_t}$ is the opinion vector at time $\pmb{t}$, $\pmb{e}$ is the vector of the interacting agents, $\pmb{s}$ encodes the interactions outcomes, and \bfeps is the latent vector ($\pmb{\varepsilon^+}$, $\pmb{\varepsilon^-}$) of ABM parameters.}

\label{fig:pgm}
\end{center}
\end{figure}